\documentclass[aps,prb,amsmath,onecolumn,superscriptaddress,floatfix,footinbib,longbibliography]{revtex4}

\usepackage{graphicx}
\usepackage{epstopdf}

\usepackage[T1]{fontenc}
\usepackage[applemac]{inputenc}
\usepackage{lmodern}
\usepackage[english]{babel}
\usepackage{ae}
\usepackage{units}

\usepackage{amsmath,amssymb,natbib,bm}
\usepackage{psfrag}
\usepackage{subfigure}
\usepackage{amsthm}

\usepackage{fixmath}
\usepackage{booktabs}

\usepackage{slashed}

\usepackage[americaninductors]{circuitikz}
\usepackage{tikz}
\usetikzlibrary{arrows}

\usepackage{url}

\usepackage[colorlinks]{hyperref}

\usepackage{tabularx}

\hypersetup{%
        plainpages=true,
        breaklinks=true,
        hypertexnames=false,
        pageanchor=true,
        colorlinks=true,
        linkcolor={blue},
        citecolor={magenta},
        urlcolor={blue},
        anchorcolor={black}
      }

\renewcommand{\eqref}[1]{\mbox{Eq.~(\ref{#1})}}

\newcommand{\ket}[1]{|#1\rangle}

\newcommand{\be}{\begin{equation}}
\newcommand{\ee}{\end{equation}}
\newcommand{\bea}{\begin{eqnarray}}
\newcommand{\eea}{\end{eqnarray}}

\begin{document}

\title{Interferometry of multi-level systems: rate-equation approach for a charge qu\emph{d}it}

\author{M.~P.~Liul}
\email[e-mail: ]{liul@ilt.kharkov.ua}
\affiliation{B.~Verkin Institute for Low Temperature Physics and Engineering, Kharkiv 61103, Ukraine}

\author{A.~I.~Ryzhov}
\email[e-mail: ]{ryzhov@ilt.kharkov.ua}
\affiliation{B.~Verkin Institute for Low Temperature Physics and Engineering, Kharkiv 61103, Ukraine}
\affiliation{Theoretical Quantum Physics Laboratory, Cluster for Pioneering Research, RIKEN, Wakoshi, Saitama 351-0198, Japan}

\author{S.~N.~Shevchenko}
\email[e-mail: ]{sshevchenko@ilt.kharkov.ua}
\affiliation{B.~Verkin Institute for Low Temperature Physics and Engineering, Kharkiv 61103, Ukraine}

\date{\today}

\begin{abstract}
We theoretically describe a driven two-electron four-level double-quantum dot (DQD) tunnel coupled to a fermionic sea by using the rate-equation formalism. This approach allows to find occupation probabilities of each DQD energy level in a relatively simple way, compared to other methods. Calculated dependencies are compared with the experimental results. The system under study is irradiated by a strong driving signal and as a result, one can observe Landau-Zener-St\"{u}ckelberg-Majorana (LZSM) interferometry patterns which are successfully described by the considered formalism. The system operation regime depends on the amplitude of the excitation signal and the energy detuning, therefore, one can transfer the system to the necessary quantum state in the most efficient way by setting these parameters. Obtained results give insights about initializing, characterizing, and controlling the quantum system states. 

\end{abstract}

\maketitle

\section{Introduction}
Fast development of the quantum computing field requires sophisticated practical solutions. One of such solutions are quantum dots. These systems are good candidates for being building blocks of quantum computers, since they have good tunability [\onlinecite{Veldhorst2014}] and flexible coupling geometry [\onlinecite{Shulman2012}]. Also, quantum dots demonstrate good performance for readout, manipulation, and initialization of their spin states [\onlinecite{Petta2005, Hanson2005, Kiyama2016, Friesen2004}]. Such systems can be used for quantum information [\onlinecite{Cerletti2005}, \onlinecite{Loss1998}] and quantum computing [\onlinecite{DiVincenzo2000, Byrd2002}]. Considered objects are also interesting for studying quantum luminescence [\onlinecite{Kim1998, Romero2009}], superconductivity [\onlinecite{Saldana2018}], Kondo effect [\onlinecite{Sasaki2004}], solid-state energy conversion [\onlinecite{Beenakker1992}], quantum communication [\onlinecite{Simon2007, Huwer2017}], piezomagnetic effect [\onlinecite{Abolfath2008}], etc.  

For solving many modern problems (one of them is a creation of a quantum computer), it is not enough to use a single quantum dot, thus one should connect them into chains [\onlinecite{Flentje2017}]. The behavior of electrons in a chain can be decomposed on interactions between pairs of adjacent dots, called double quantum dots (DQDs). As a result these systems are widely explored nowadays. Particularly, it was stressed that DQDs open opportunities for probing electron-phonon coupling [\onlinecite{Brandes2015}], allow to probe the semiconductor environment [\onlinecite{VanderWiel2003}], can be used in the relatively new and promising area of spintronics [\onlinecite{Awschalom2007}], serve as thermoelectric generators [\onlinecite{Donsa2014}] and noise detectors [\onlinecite{Aguado2000}]. Therefore, both experimental and theoretical study of such systems is very important not only from quantum information point of view but also for modern quantum physics in general. 

In the current paper we theoretically study the properties of a qu\emph{d}it (\emph{d}-level quantum system) [\onlinecite{Liu_2017, Wang2020, Han2019, Kononenko2021, Yurtalan2020}] similar to the one experimentally studied in Ref.~[\onlinecite{Chatterjee2018}]. The main tool of our analysis is the rate-equation formalism [\onlinecite{Ferron2012, Ferron2016, Liul2022}], which is relatively simple, but often shows good agreement with experiments. For example, this method successfully describes the behavior of a two-level system [\onlinecite{Berns2006}] as well as a multi-level system (solid-state artificial atom) from Ref.~[\onlinecite{Oliver2009}] which was analyzed in Ref.~[\onlinecite{Wen2009}].

Present research could also be interesting since it opens an additional opportunity for studying the Landau-Zener-St\"{u}ckelberg-Majorana (LZSM) transitions. This effect can be observed if one irradiates a quantum system by a signal with the frequency which is much smaller than the distance between energy levels [\onlinecite{Izmalkov2004, Ashhab2014}]. LZSM transitions appear in many fields; for instance, in solid-state physics [\onlinecite{Nakonechnyi2021, Chien2023}], quantum information science [\onlinecite{Fuchs2011, Ribeiro2009}], nuclear physics [\onlinecite{Thiel1990}], chemical physics [\onlinecite{Zhu1997}], and quantum optics [\onlinecite{Bouwmeester1995}]. Repeated LZSM transitions result in LZSM interference [\onlinecite{Stehlik2012, Oliver2005, Gonzalez_Zalba_2016, Kofman2022, Ono2019}]. The LZSM interferometry can be used for a quantum system description and control [\onlinecite{Wu19, Ivakhnenko2022}]; it allows to understand better processes of photon-assisted transport in superconducting systems [\onlinecite{Nakamura1999}] and decoherence in quantum systems [\onlinecite{Rudner2008, Malla2022}]. 

The rest of the paper is organized as follows. In Sec.~II the rate-equation formalism for a TLS is laid out with its subsequent generalization on multi-level systems. The considered model is applied for the analysis of the two-electron double quantum dot in Sec.~III. Section~IV is devoted to a DQD studied in the three-level approximation. In Sec.~V we present our conclusions. The expressions for building DQD energy levels diagram are presented in Appendix~A.
\begin{figure*}
	\includegraphics[width=1 \linewidth]{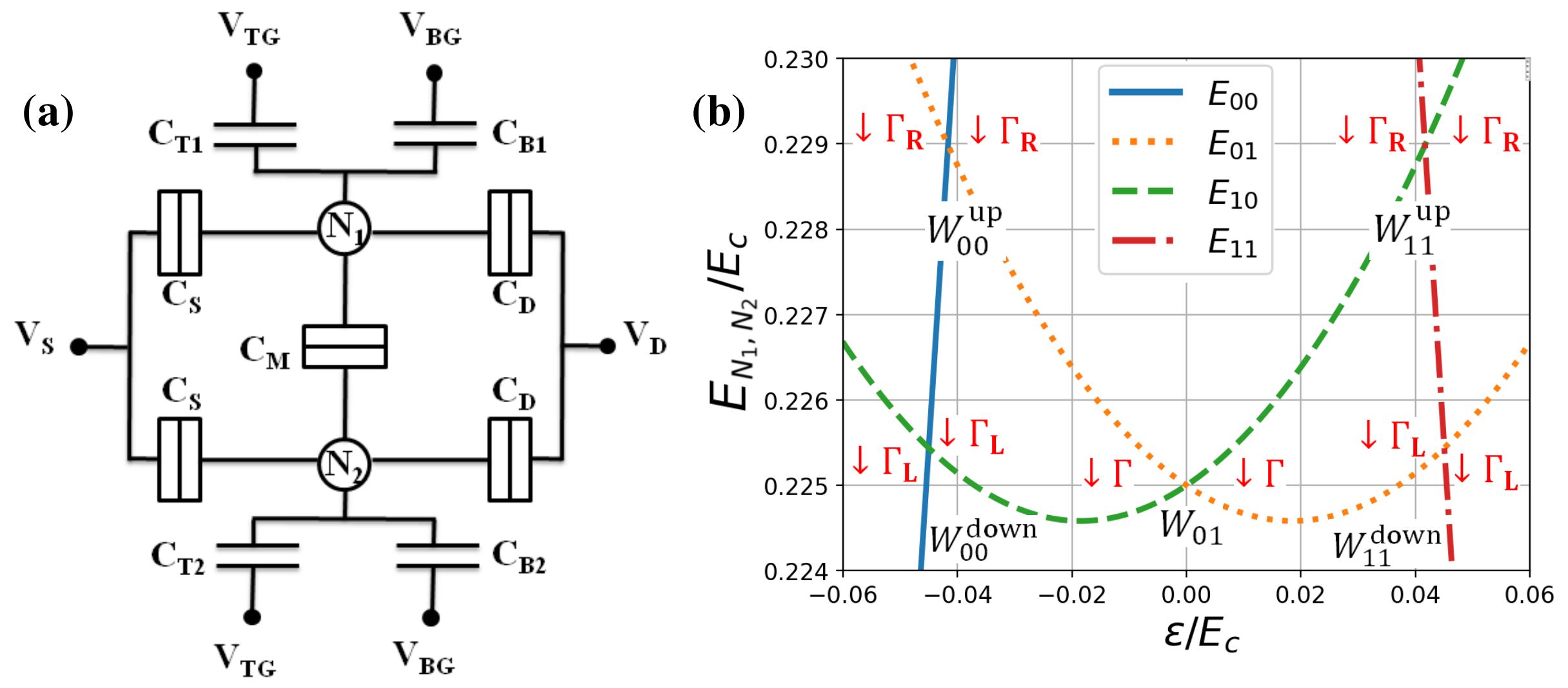}
	\caption{Scheme of the DQD and its energy levels in the charge basis. Panel (a) shows the electrical scheme of the considered DQD, with $N_{1}$ and $N_{2}$ being the numbers of electrons in each dot, $V$ and $C$ with the subscripts indicate the respective applied voltages and capacitances. (b) Energy levels of the DQD similar to the one studied in Ref.~[\onlinecite{Chatterjee2018}] in the charge basis. The expression for the electrostatic energy $E_{N_1, N_2}$ is given by Eq.~(\ref{eq:U}), the characteristic charging energy $E_{C}$ is defined after Eq.~(\ref{eq:U}). The values $W_{00}^{\rm up}$ and $W_{00}^{\rm down}$ are the transition rates between the state $\left | 00 \right \rangle $ and the states $\left | 01 \right \rangle$, $\left | 10 \right \rangle$, respectively. The values $W_{11}^{\rm up}$ and $W_{11}^{\rm down}$ are the transition rates between the state $\left | 11 \right \rangle $ and the states $\left | 10 \right \rangle$, $\left | 01 \right \rangle$, respectively. The transition rate $W_{01}$ indicates transitions between the states $\left | 01 \right \rangle$ and $\left | 10 \right \rangle$. The value $\Gamma_{\rm L}$ is the relaxation rate between the states $\left | 00 \right \rangle$ and $\left | 10 \right \rangle$, $\left | 11 \right \rangle$ and $\left | 01 \right \rangle$ (left dot and reservoirs in terms of the experiment). The value $\Gamma_{\rm R}$ is the relaxation rate between the states $\left | 00 \right \rangle$ and $\left | 01 \right \rangle$, $\left | 11 \right \rangle$ and $\left | 10 \right \rangle$  (right dot and reservoirs in terms of the experiment), $\Gamma$ is the relaxation rate between states $\left | 01 \right \rangle$ and $\left | 10 \right \rangle$.}
		\label{energy_levels}
\end{figure*}

\section{Rate-equation approach}
In this section, we describe theoretical aspects of the rate-equation formalism. For doing this, we first employ this method for a two-level system with further extension of obtained results on multi-level systems. The Hamiltonian of a TLS driven by an external field can be written in the form
\begin{eqnarray}
 \widehat{H}(t) = -\frac{\mathrm{\Delta}}{2}\widehat{\sigma}_{x} - \frac{h(t)}{2}\widehat{\sigma}_{z}, 
\label{eq:Hamiltonian_TLS}
\end{eqnarray}
where $\widehat{\sigma}_{z} = \begin{pmatrix} 1 & 0 \\ 0 & -1 \end{pmatrix}$ and $\widehat{\sigma}_{x} = \begin{pmatrix} 0 & 1 \\ 1 & 0 \end{pmatrix}$ are Pauli matrices, $\mathrm{\Delta}$ is the level splitting, $h(t)$ is the external excitation which can be presented as follows:
\begin{eqnarray}
 h(t) = \varepsilon + A\sin2\pi\nu t + \delta \varepsilon_{\mathrm{noise}}(t).
\label{eq:excitation}
\end{eqnarray}
Here, $\varepsilon$ is an energy detuning, $\nu$ and $A$ are the frequency of the excitation field and its amplitude, respectively, $\delta\varepsilon_{\mathrm{noise}}(t)$ can be treated as the classical noise. In Ref.~[\onlinecite{Berns2006}] the authors used the white-noise model and for the LZSM transition rate they obtained (see also Refs.~[\onlinecite{Chen2011, Wang2010, Wen2010, Otxoa2019, He2023}])
\begin{eqnarray}
W(\varepsilon, A) = \frac{\mathrm{\Delta}^{2}}{2}\sum_{n}\frac{\Gamma_{2}J_{n}^{2}(A/\nu)}{\left ( \varepsilon - n \nu \right )^{2} + \Gamma_{2}^{2}}.
\label{eq:transition_rate}
\end{eqnarray}
Here, $\Gamma_{2}$ is the decoherence rate, $J_{n}$ is the Bessel function, and the reduced Planck constant is equal to unity ($\hbar~=~1$). Equation~(\ref{eq:transition_rate}) characterizes the transitions which happen when a system passes through a point of maximum levels approaching.

In the case of a multi-level system we should assign a corresponding transition rate to each level quasicrossing point (point of maximum levels approaching). 
The authors of Ref.~[\onlinecite{Wen2009}] proposed to extend Eq.~(\ref{eq:transition_rate}) on the transition between arbitrary states $\ket{i}$ and $\ket{j}$ of a multi-level system by the formula: 
\begin{eqnarray}
W_{ij}(\varepsilon_{ij}, A) = \frac{\mathrm{\Delta}_{ij}^{2}}{2}\sum_{n}\frac{\Gamma_{2}J_{n}^{2}(A/\nu)}{\left ( \varepsilon_{ij} - n \nu \right )^{2} + \Gamma_{2}^{2}},
\label{eq:transition_rate_general}
\end{eqnarray}
where $\mathrm{\Delta}_{ij}$ is the energy splitting between states $\ket{i}$ and $\ket{j}$, $\varepsilon_{ij}$ is the corresponding energy detuning. The validity of the generalization of Eq.~(\ref{eq:transition_rate}) to Eq.~(\ref{eq:transition_rate_general}) is an empirical approach. Here, we refer to Refs.~[\onlinecite{Wen2009, Liul2022, Oliver2009}], where this is discussed in more detail. At this point, we would like to explain that for our system and for the one from Ref.~[\onlinecite{Oliver2009}] (the theoretical description was done in Refs.~[\onlinecite{Wen2009, Liul2022}]), this approach gives good quantitative results, which was confirmed by comparison with both experimental observations and numerical calculations. Particularly for Refs.~[\onlinecite{Wen2009, Liul2022, Oliver2009}], the structure of the first two diamonds in the theory and in the experiment shows the same patterns: for the first diamond, the multiphoton resonances are not distinguishable, while for the second one, they are separated. The form and the positions of resonances are in a good agreement as well. For the case of our theoretical results in this paper, we can also conclude that the obtained picture shows the same patterns as the experimental ones in Ref.~[\onlinecite{Chatterjee2018}]. In particular, one can observe four different LZSM regimes (multi-passage, single-passage, double-passage, incoherent) what will be discussed further in more detail.
 
The rate equation for the $\ket{i}$ state can be expressed
\begin{eqnarray}
\frac{d P_{i}}{dt} = \sum_{j}W_{ij}(P_{j} - P_{i}) + \sum_{i'}\Gamma_{i' i}P_{i'} - \sum_{i'}\Gamma_{i i'}P_{i}. 
\label{eq:rate_equation_general}
\end{eqnarray}
Here, $P_{i}$ is the probability that a system occupies $\ket{i}$ state, $\Gamma_{ii'}$ characterize the relaxation from the state $\ket{i}$ to the state $\ket{i'}$. \\

Thus, writing equations (\ref{eq:rate_equation_general}) for each level, we can find occupation probabilities of the levels and then build corresponding interferograms. Usually, for simplicity, one considers only a stationary case, $d P_{i}/dt = 0$. The solution of such a system will not describe quantum dynamics, but it is suitable for obtaining its main properties. Also we can use the fact that the sum of all probabilities is equal to unity $\sum_{i}P_{i} = 1$. 

\section{Rate equations and interferogram for the DQD}

In this section, we apply the rate-equation formalism to the parallel DQD. The scheme of the considered system is depicted in Fig.~\ref{energy_levels}(a), where $N_{1}$ and $N_{2}$ are numbers of electrons in each dot, $V$ and $C$ with the corresponding subscripts indicate the applied voltages and capacitances, respectively. Figure~\ref{energy_levels}(b) shows the energy levels diagram. The detailed procedure of an energy levels diagram building from the scheme in Fig.~\ref{energy_levels}(a) is described in Appendix~A.

We consider charge states of the system. As a result after each point of maximum levels approaching (their positions for $\mathrm{\Delta} _{\mathrm{c}}$ is at $\varepsilon_{\mathrm{c}}~=~\unit[0]{}$, for $\mathrm{\Delta}^{\rm down}$ are at $\varepsilon_{00}^{\rm down} =~\unit[-276]{GHz}$ and $\varepsilon_{11}^{\rm down} =~\unit[276]{GHz}$, for $\mathrm{\Delta}^{\rm up}$ are at $\varepsilon_{00}^{\rm up} =~\unit[-255]{GHz}$ and $\varepsilon_{11}^{\rm up} =~\unit[255]{GHz}$) levels swap their positions (an upper level becomes a lower one). Therefore, we should take this fact into account. This effect will especially have an influence on relaxations which occur from the upper level to the lower one. We imply that inverse relaxations are Boltzmann suppressed. To handle this we split our interval into two parts: for the first interval $\varepsilon < 0$ and for the second interval $\varepsilon \geq 0$. Each of these parts is described by different systems of rate equations (the relaxation terms differ, other terms do not change). In our calculations (not shown here) we also split our picture into more parts (for example, we have split the interval into six parts as the following: (i) $\varepsilon < \varepsilon_{00}^{\rm down}$, (ii) $\varepsilon_{00}^{\rm down} \leq \varepsilon < \varepsilon_{00}^{\rm up}$, (iii) $\varepsilon_{00}^{\rm up} \leq \varepsilon < \varepsilon_{\mathrm{c}}$, (iv) $\varepsilon_{\mathrm{c}} \leq \varepsilon < \varepsilon_{11}^{\rm up}$, (v) $\varepsilon_{11}^{\rm up} \leq \varepsilon < \varepsilon_{11}^{\rm down}$, (vi) $\varepsilon \geq \varepsilon_{11}^{\rm down}$), but it had worse performance. Therefore, finally, for the considered case the rate equations (\ref{eq:rate_equation_general}) take the following form

Interval I ($\varepsilon < 0$):
\begin{align}
\resizebox{0.75\textwidth}{!}{$
\begin{cases}
\dot{P_{01}} = W_{10}\left (P_{10} - P_{01} \right ) + W_{11}^{\rm down}\left ( P_{11} - P_{01} \right ) + W_{00}^{\rm up}\left (P_{00} - P_{01} \right ) - \Gamma P_{01} + \Gamma_{\rm R}P_{00} + \Gamma_{\rm L} P_{11} \\ 
\dot{P_{10}} = W_{11}^{\rm up}\left (P_{11} - P_{10} \right ) + W_{00}^{\rm down}\left ( P_{00} - P_{10} \right ) + W_{10}\left ( P_{01} - P_{10} \right )  + \Gamma P_{01} + \Gamma_{\rm L} P_{00} + \Gamma_{\rm R} P_{11} \\ 
\dot{P_{00}} = W_{00}^{\rm down}\left ( P_{10} - P_{00} \right ) + W_{00}^{\rm up}\left ( P_{01} - P_{00} \right ) - \Gamma_{\rm R}P_{00} - \Gamma_{\rm L} P_{00} \\
P_{01} + P_{10} + P_{00} + P_{11} = 1\\
\end{cases}$}
\label{eq:four_level_system_1}
\end{align}
Interval II ($\varepsilon \geq 0$):
\begin{align}
\resizebox{0.75\textwidth}{!}{$
\begin{cases}
\dot{P_{01}} = W_{10}\left (P_{10} - P_{01} \right ) + W_{11}^{\rm down}\left ( P_{11} - P_{01} \right ) + W_{00}^{\rm up}\left (P_{00} - P_{01} \right ) + \Gamma P_{10} + \Gamma_{\rm R}P_{00} + \Gamma_{\rm L} P_{11} \\ 
\dot{P_{10}} = W_{11}^{\rm up}\left (P_{11} - P_{10} \right ) + W_{00}^{\rm down}\left ( P_{00} - P_{10} \right ) + W_{10}\left ( P_{01} - P_{10} \right )  - \Gamma P_{10} + \Gamma_{\rm L} P_{00} + \Gamma_{\rm R} P_{11}\\ 
\dot{P_{00}} = W_{00}^{\rm down}\left ( P_{10} - P_{00} \right ) + W_{00}^{\rm up}\left ( P_{01} - P_{00} \right ) - \Gamma_{\rm R}P_{00} - \Gamma_{\rm L} P_{00} \\
P_{01} + P_{10} + P_{00} + P_{11} = 1\\
\end{cases}$}
\label{eq:four_level_system_2}
\end{align}

The transition rates can be calculated by the following expressions
\begin{eqnarray}
W_{10} &=&\frac{\mathrm{\Delta} _{\mathrm{c}}^{2}}{2}\sum_{n}\frac{\Gamma
_{2}J_{n}^{2}\left( \frac{A}{\nu }\right) }{\left( 
(\varepsilon - \varepsilon_{\mathrm{c}}) -n\nu \right) ^{2}+ \Gamma _{2}^{2}}, \notag \\
W_{00}^{\rm up } &=&\frac{(\mathrm{\Delta}^{\rm up})^{2}}{2}\sum_{n}\frac{\Gamma
_{2,\rm R}J_{n}^{2}\left( \frac{A}{\nu }\right) }{\left((\varepsilon - \varepsilon_{00}^{\rm up}) - n\nu \right) ^{2}+ \Gamma _{2,\rm R}^{2}}, \notag
\\
W_{00}^{\rm down} &=&\frac{(\mathrm{\Delta}^{\rm down})^{2}}{2}\sum_{n}\frac{\Gamma
_{2,\rm L}J_{n}^{2}\left(\frac{A}{\nu }\right) }{\left((\varepsilon - \varepsilon_{00}^{\rm down}) - n\nu \right) ^{2}+ \Gamma _{2,\rm L}^{2}}, \notag
\\
W_{11}^{\rm up } &=&\frac{(\mathrm{\Delta}^{\rm up})^{2}}{2}\sum_{n}\frac{\Gamma
_{2,\rm R}J_{n}^{2}\left( \frac{A}{\nu }\right) }{\left((\varepsilon - \varepsilon_{11}^{\rm up}) - n\nu \right) ^{2}+ \Gamma _{2,\rm R}^{2}}, \notag
\\
W_{11}^{\rm down} &=&\frac{(\mathrm{\Delta}^{\rm down})^{2}}{2}\sum_{n}\frac{\Gamma
_{2,\rm L}J_{n}^{2}\left(\frac{A}{\nu }\right) }{\left((\varepsilon - \varepsilon_{11}^{\rm down}) - n\nu \right) ^{2}+ \Gamma _{2,\rm L}^{2}}, \notag
\end{eqnarray}
where $\Gamma_{2}~=~\unit[4]{GHz}$. Also, we take into account the relation between the relaxation rate $\Gamma_{1}$ and the decoherence rate $\Gamma_{2}$ of the system $\Gamma_{2} \approx \Gamma_{1}/2$ (in general case, this relation can be written as $\Gamma_{2} \approx \Gamma_{1}/2 + \Gamma_{\phi}$, where $\Gamma_{\phi}$ is a system dephasing rate, but in our case, we neglect this term). Then,  $\Gamma _{2,\rm R} \approx \Gamma _{\rm R} / 2$, $\Gamma _{2,\rm L} \approx \Gamma _{\rm L} / 2$. For numerics, we used the following parameters: relaxation rates of the system are $\Gamma_{1}~=~\unit[0.9]{GHz}$, $\Gamma_{\rm R}~=~\unit[12]{GHz}$, $\Gamma_{\rm L}~=~\unit[50]{MHz}$, the energy splittings are equal to $\mathrm{\Delta}_{\rm c}~=~\unit[8.25]{GHz}$,  $\mathrm{\Delta}^{\rm down}~=~\unit[21]{GHz}$, $\mathrm{\Delta}^{\rm up}~=~\unit[1]{GHz}$, and their positions are at $\varepsilon_{\mathrm{c}}~=~\unit[0]{}$,  $\varepsilon_{00}^{\rm down} =~\unit[-276]{GHz}$,  $\varepsilon_{11}^{\rm down} =~\unit[276]{GHz}$, $\varepsilon_{00}^{\rm up} =~\unit[-255]{GHz}$,  $\varepsilon_{11}^{\rm up} =~\unit[255]{GHz}$, respectively. The decoherence rate $\Gamma_{2}~=~\unit[4]{GHz}$ and the excitation frequency $\nu~=~\unit[4]{GHz}$.

\begin{figure*}
	\includegraphics[width=1 \linewidth]{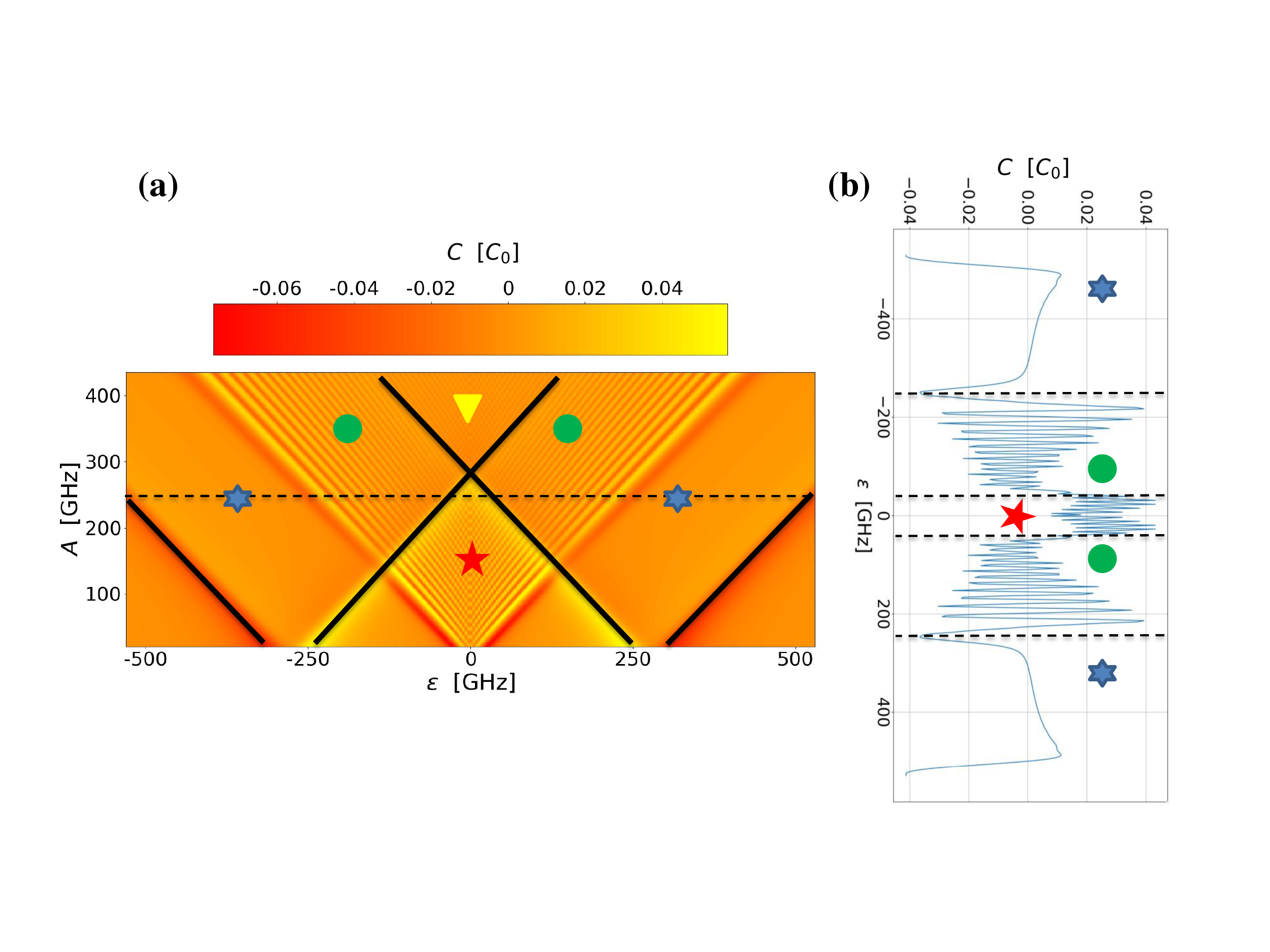}
	\caption{(a) Parametric capacitance $C$ of the DQD as a function of the excitation field amplitude $A$ and the energy detuning $\varepsilon$. This value is calculated in units of $C_{0}$. (b) Dependence of the parametric capacitance of the DQD on the energy detuning $\varepsilon$ for the excitation field amplitude $A ~=~\unit[240]{GHz}$ (a line cut of panel (a) along $y$-axis). The corresponding relaxation rates of the system are $\Gamma_{1}~=~\unit[0.9]{GHz}$, $\Gamma_{\rm R}~=~\unit[12]{GHz}$, $\Gamma_{\rm L}~=~\unit[50]{MHz}$. The energy splittings are equal to $\mathrm{\Delta}_{\rm c}~=~\unit[8.25]{GHz}$, $\mathrm{\Delta}^{\rm down}~=~\unit[21]{GHz}$, $\mathrm{\Delta}^{\rm up}~=~\unit[1]{GHz}$, and their positions are at $\varepsilon_{\mathrm{c}}~=~\unit[0]{}$,  $\varepsilon_{00}^{\rm down} =~\unit[-276]{GHz}$,  $\varepsilon_{11}^{\rm down} =~\unit[276]{GHz}$, $\varepsilon_{00}^{\rm up} =~\unit[-255]{GHz}$,  $\varepsilon_{11}^{\rm up} =~\unit[255]{GHz}$. The decoherence rate $\Gamma_{2}~=~\unit[4]{GHz}$ and the excitation frequency $\nu~=~\unit[4]{GHz}$.} 
		\label{interferogram}
\end{figure*}

\begin{figure*}
	\includegraphics[width=1 \linewidth]{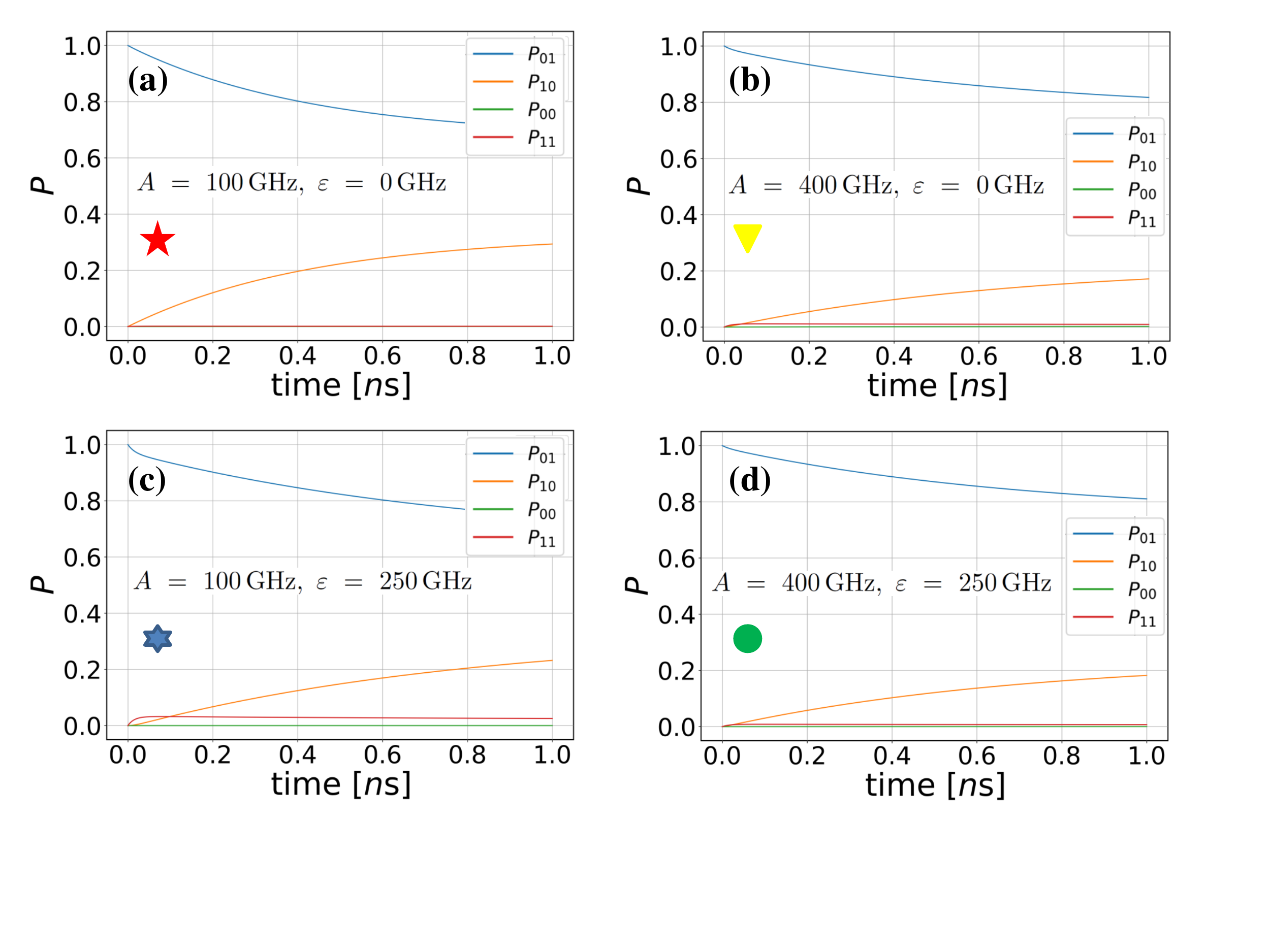}
	\caption{Time dependence of probabilities for different regimes: (a) multi-passage LZSM; (b) single-passage LZSM; (c) incoherent LZSM; (d) double-passage LZSM. The parameters are the same with Fig.~\ref{interferogram}.} 
		\label{time}
\end{figure*}

By solving Eqs.~(\ref{eq:four_level_system_1}, \ref{eq:four_level_system_2}) for the stationary regime (when $dP_{ij}/dt = 0$), we obtain $P_{mn} = P_{mn}(\varepsilon,A)$, $m, n=0,1$, as a function of $\varepsilon$ and $A$ what allows us to analyze the experimentally measured value, the phase response of the resonator $\mathrm{\Delta} \phi$, which according to Ref.~[\onlinecite{Chatterjee2018}] can be written as
\begin{equation}
\mathrm{\Delta} \phi = -\pi Q \frac{C}{C_{\rm p}},
\end{equation}
where $C_{\rm p}~=~\unit[660]{fF}$ is the parasitic capacitance to ground of the device, $Q = 42$ is the loaded Q-factor of the resonator, $C$ is the parametric capacitance of a DQD which can be calculated by the following expression
\begin{equation}
C~=~C_{0}\frac{d}{d\varepsilon}\left\{ P_{01}-P_{10}+\alpha\left( P_{00}-P_{11}\right) \right\},
\end{equation}
where $\alpha$ is a dimensionless factor which describes the DQD coupling to the fermionic sea, for the experiment of Ref.~[\onlinecite{Chatterjee2018}] $\alpha~=~18$ and $C_{0}$ is the constant proportionality factor. In our calculations we plot the parametric capacitance $C$ as a function of $\varepsilon$, and $A$. The results of the theoretical calculations are presented in Fig.~\ref{interferogram}(a). The obtained interferogram shows the patterns similar to the experimental ones (see Fig.~3(b) in Ref.~[\onlinecite{Chatterjee2018}]). Specifically, one can see four different regimes: incoherent one (blue star), double-passage LZSM (green circle), single-passage LZSM (yellow triangle), multi-passage LZSM (red star). Fig.~\ref{interferogram}(b) presents a line cut of Fig.~\ref{interferogram}(a) at $A ~=~\unit[240]{GHz}$, the same patterns can be seen in the experiment (see Fig.~3(c) in Ref.~[\onlinecite{Chatterjee2018}]). 

Figure~\ref{time} shows time dependence of probabilities for different regimes: (a) multi-passage LZSM; (b) single-passage LZSM; (c) incoherent one; (d) double-passage LZSM. From the plots, one can conclude that the stationary regime (when $dP_{ij}/dt~\approx~0$) starts after $t_{0}~\approx~\unit[0.8]{ns}$. Also, it could be seen that for all cases $P_{01}>P_{10}\gg P_{11}, P_{00}$.

\section{Three-level approximation of the DQD}
\begin{figure*}
	\includegraphics[width=0.65 \linewidth]{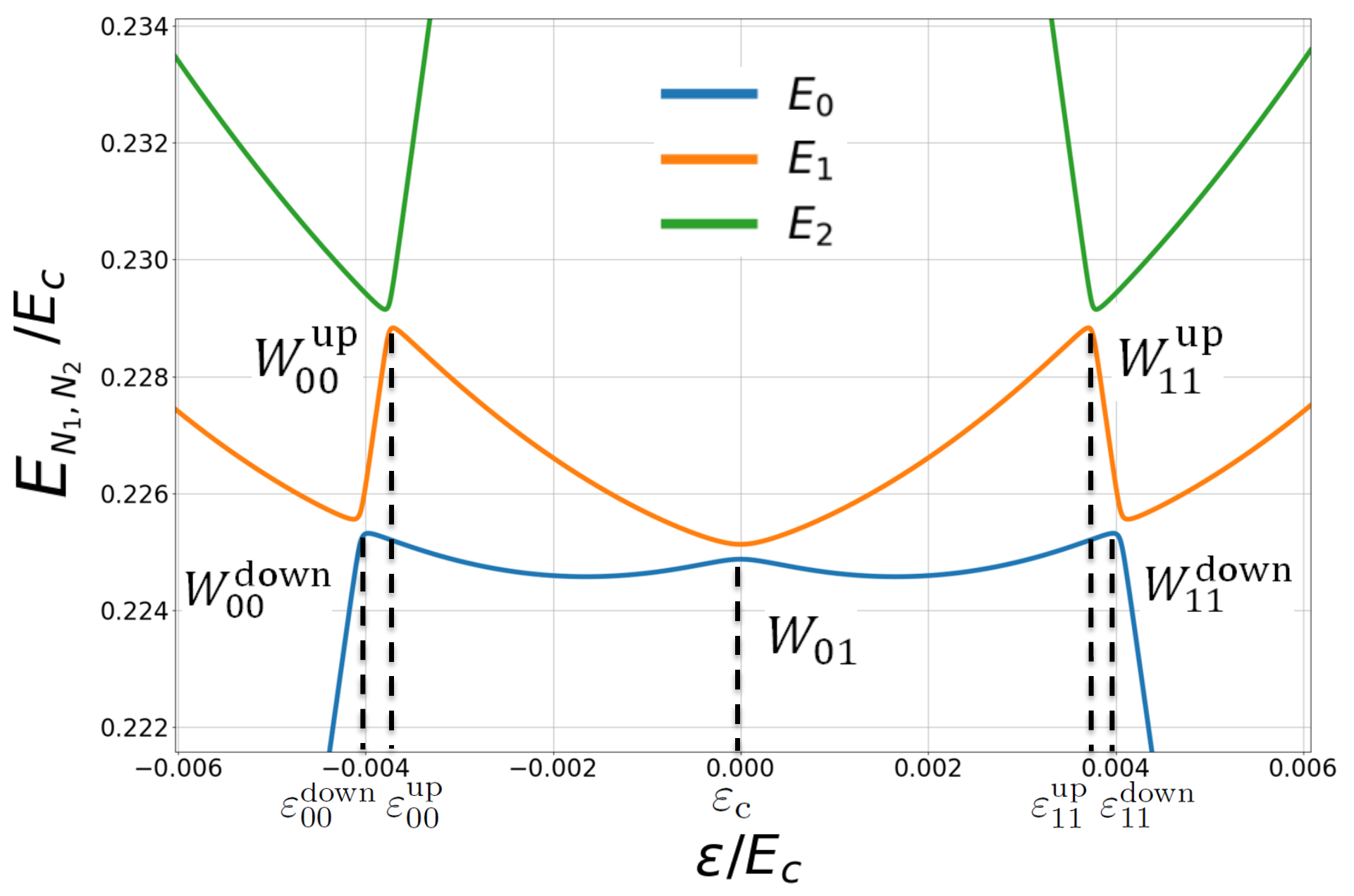}
	\caption{Energy levels of the DQD in the energy basis. This plot is a representation of Fig.~\ref{energy_levels}(b) in the energy basis. The value $W_{01}$ indicates transition rate between energy states $E_1$ and $E_0$ at $\varepsilon~=~\varepsilon_{\rm c}$, $W_{00}^{\rm down}$ and $W_{11}^{\rm down}$ indicate transition rates between energy states $E_1$ and $E_0$ at $\varepsilon~=~\varepsilon_{00}^{\rm down}$ and $\varepsilon_{11}^{\rm down}$, respectively, $W_{00}^{\rm up}$ and $W_{11}^{\rm up}$ indicate transition rates between energy states $E_1$ and $E_2$ at $\varepsilon~=~\varepsilon_{00}^{\rm up}$ and $\varepsilon_{11}^{\rm up}$, respectively.}
		\label{energy_levels_e_basis}
\end{figure*}

In this section, we consider the system in the energy basis (in the previous section, the system was described in the charge basis) with the basis states $E_0,~E_1,~E_2,~E_3$ and probabilities to occupy the corresponding energy state $P_{0},~P_{1},~P_{2},~P_{3}$. The corresponding energy-level diagram is shown in Fig.~\ref{energy_levels_e_basis}. In the considered case, we can neglect the highest energy level ($P_{3}=0$) and take into account only three levels. Our energy-level diagram has five avoided level crossings at $\varepsilon_{\mathrm{c}}~=~\unit[0]{}$,  $\varepsilon_{00}^{\rm down} =~\unit[-276]{GHz}$,  $\varepsilon_{11}^{\rm down} =~\unit[276]{GHz}$, $\varepsilon_{00}^{\rm up} =~\unit[-255]{GHz}$,  $\varepsilon_{11}^{\rm up} =~\unit[255]{GHz}$. 

The rate equations (\ref{eq:rate_equation_general}) take the form%
\begin{equation}
\left\{ 
\begin{array}{c}
\frac{dP_{0}}{dt}=\left[ W_{\mathrm{01}}+W_{00}^{\rm down}+W_{11}^{\rm down}\right] \left(
P_{1}-P_{0}\right) +\Gamma _{1\rightarrow 0}P_{1}, \\ 
\frac{dP_{1}}{dt}=\left[ W_{\mathrm{01}}+W_{00}^{\rm down}+W_{11}^{\rm down}\right] \left(
P_{0}-P_{1}\right) + \\ 
+\left[ W_{00}^{\rm up}+W_{11}^{\rm up}\right] \left( P_{2}-P_{1}\right) -\Gamma
_{1\rightarrow 0}P_{1}, \\ 
P_{0}+P_{1}+P_{2}=1.%
\end{array}%
\right.  \label{System}
\end{equation}%

These equations contain only leading terms; in particular, we omitted the term $\Gamma _{2\rightarrow 0}P_{2}$ in the first equation and the term $%
\Gamma _{2\rightarrow 1}P_{2}$ in the second equation. The fact that there
are different relaxations between different levels is taken into account by
assuming the time-dependent relaxation $\Gamma _{1\rightarrow 0}=\Gamma
_{1\rightarrow 0}(t)$:%
\begin{equation}
\Gamma _{1\rightarrow 0}(t)=\left\{ 
\begin{array}{c}
\Gamma _{1},\text{ \ }\left\vert \varepsilon (t)\right\vert <\varepsilon_{00}^{\rm down}, \\ 
\Gamma _{\mathrm{L}},\text{ }\ \left\vert \varepsilon (t)\right\vert
>\varepsilon_{00}^{\rm down},%
\end{array}%
\right.  \label{Gamma10}
\end{equation}%
where the first line describes the inter-dot relaxation, while the second
line corresponds to the tunneling between the left dot and the leads.

Relation (\ref{Gamma10}) provides two simplifications, when the whole
dynamics is either for $\left\vert \varepsilon (t)\right\vert < \varepsilon_{00}^{\rm up}$ or for $\ \left\vert \varepsilon (t)\right\vert >\varepsilon_{00}^{\rm up}$. In these cases $\Gamma _{1\rightarrow 0}$ becomes
independent of time and we can have the stationary solution of the system of
equations~(\ref{System}). (Then we can have the recipe for alike
situations: take the analytical solutions for such regions, then fitting
with this is a simple way to get the parameters; and afterward, one can
continue with more elaborated calculations, such as solving time-dependent
equations.) In these two particular cases we obtain analytical stationary
solutions. For the coherent regime (\textquotedblleft
red-star\textquotedblright \ region) we have $W_{00}^{\rm down}$, $W_{00}^{\rm up}$, $W_{11}^{\rm down}$ and $W_{11}^{\rm up}$ $\rightarrow 0$ and 
$P_{2}\rightarrow 0$, and then, it follows [\onlinecite{Berns2006, Ivakhnenko2022}]

\begin{eqnarray}
P_{1} &=&\frac{1}{2}\sum_{n=-\infty }^{\infty }\frac{\mathrm{\Delta} _{\mathrm{c}%
,n}^{2}}{\mathrm{\Delta} _{\mathrm{c},n}^{2}+\frac{\Gamma _{1}}{\Gamma _{2}}\left( 
(\varepsilon - \varepsilon_{\mathrm{c}})-n\nu \right) ^{2}+\Gamma _{1}\Gamma _{2}},
\label{P1} \\
\mathrm{\Delta} _{\mathrm{c},n} &=&\mathrm{\Delta} _{\mathrm{c}}J_{n}\left( \frac{A}{\nu}\right).  \notag
\end{eqnarray}%
Analogously, for the incoherent regime (\textquotedblleft
blue-star\textquotedblright\ region) nearby the point $\varepsilon =\varepsilon_{00}^{\mathrm{down}}$  we have $W_{01},W_{00}^{\rm up}, W_{11}^{\rm up}$ and $W_{11}^{\rm down}\rightarrow 0$ and $P_{2}\rightarrow 0$, therefore, then it follows

\begin{eqnarray}
P_{1} &=&\frac{1}{2}\sum_{n=-\infty }^{\infty }\frac{(\mathrm{\Delta}_{n}^{\rm down})^{2}}{(\mathrm{\Delta}_{n}^{\rm down})^{2}+\frac{\Gamma _{\mathrm{L}}}{\Gamma
_{2,L}}\left( (\varepsilon - \varepsilon_{00}^{\mathrm{down}})-n\nu \right)
^{2}+\Gamma _{\mathrm{L}}\Gamma _{2,L}}, \\
\mathrm{\Delta}_{n}^{\rm down} &=&\mathrm{\Delta}^{\rm down}J_{n}\left( \frac{A}{
\nu }\right) .  \notag
\end{eqnarray}%
We can also note that in the lower parts of the $(\varepsilon, A)$ plane,
below the red-star and the blue-star regions, all $W$'s are $0$, and we have $P_{i}$ being all constants, resulting in zero $C$.

\section{Conclusions}
We considered the rate-equation approach for a theoretical description of the four-level DQD. The system state as a function of the energy detuning $\varepsilon$  and the amplitude of the excitation signal $A$ was studied. We obtained that the DQD can be operated in four regimes in dependence on the considered parameters. These regimes are single-passage LZSM, which corresponds to small $\varepsilon$ and large $A$; double-passage LZSM (large $A$ and medium $\varepsilon$); multi-passage LZSM (small $A$ and $\varepsilon$); incoherent regime (large $\varepsilon$ and small $A$). Our research gives information about the DQD properties and behavior which could be used for the system initializing and controlling.

\begin{acknowledgments}
S.N.S. acknowledges fruitful discussions with M.F.~Gonzalez-Zalba and Franco Nori. M.P.L. was partially supported by the grant from the National Academy of Sciences of Ukraine for research works of young scientists. A.I.R. was supported by the RIKEN International Program Associates (IPA). This work was supported by Army Research Office (ARO) (Grant No.~W911NF2010261). 

\end{acknowledgments}
\appendix
\section{Energy levels of a parallel double-quantum dot}

In the current research we study the DQD proposed in Ref.~[\onlinecite{Chatterjee2018}],
see Fig.~\ref{energy_levels}(a). The first step in the system analysis is finding of the DQD energy levels. In the general form system electrostatic energy can be written by 
\begin{equation}
E=\frac{1}{2}\overrightarrow{V}\cdot \mathbf{C}\overrightarrow{V}=\frac{1}{2}%
\overrightarrow{V}\cdot \overrightarrow{Q}=\frac{1}{2}\overrightarrow{Q}%
\cdot \mathbf{C}^{-1}\overrightarrow{Q},  \label{eq:energy_levels_general}
\end{equation}%
where $\overrightarrow{V}$ and $\overrightarrow{Q}$ are the vectors of voltages and charges, respectively, $\mathbf{C}$ is the capacitance matrix.
In Eq.~(\ref{eq:energy_levels_general}) we used $\overrightarrow{Q}=\mathbf{C%
}\overrightarrow{V}$. Thus, to obtain the system energy levels, we need to
find the vector of charges and the inverse capacitance matrix of the system.%

The charges $Q_{1,2}$ in the quantum dots can be written as follows 
\begin{eqnarray}
Q_{1} &=&C_{\rm T1}(V_{1}-V_{\rm TG})+C_{\rm B1}(V_{1}-V_{\rm BG})+C_{\rm S}(V_{1}-V_{\rm S})+ 
\notag \\
&&+C_{\rm D}(V_{1}-V_{\rm D})+C_{\rm M}(V_{1}-V_{2}),  \label{eq:Charge_in_dot1}
\end{eqnarray}

\begin{eqnarray}
Q_{2} &=&C_{\rm T2}(V_{2}-V_{\rm TG})+C_{\rm B2}(V_{2}-V_{\rm BG})+C_{\rm S}(V_{2}-V_{\rm S})+ 
\notag \\
&&+C_{\rm D}(V_{2}-V_{\rm D})+C_{\rm M}(V_{2}-V_{1}).  \label{eq:Charge_in_dot2}
\end{eqnarray}%
It is convenient to rewrite expressions (\ref{eq:Charge_in_dot1}, \ref%
{eq:Charge_in_dot2}) in a matrix form

\begin{eqnarray}
\overrightarrow{Q} &=&%
\begin{pmatrix}
Q_{1}+C_{\rm T1}V_{\rm TG}+C_{\rm B1}V_{\rm BG}+C_{\rm S}V_{\rm S}+C_{\rm D}V_{\rm D} \\ 
Q_{2}+C_{\rm T2}V_{\rm TG}+C_{\rm B2}V_{\rm BG}+C_{\rm S}V_{\rm S}+C_{\rm D}V_{\rm D}%
\end{pmatrix}%
=  \notag \\
&&%
\begin{pmatrix}
C_{1}V_{1}-C_{\rm M}V_{2} \\ 
C_{2}V_{2}-C_{\rm M}V_{1}%
\end{pmatrix}%
,  \label{eq:charge_matrix}
\end{eqnarray}%
where $C_{1}$ and $C_{2}$ are the capacitances, connected to the first and
the second quantum dots, respectively,

\begin{eqnarray}
C_{1} &=&C_{\rm T1}+C_{\rm B1}+C_{\rm D}+C_{\rm S}+C_{\rm M},  \label{eq:expr_C1} \\
C_{2} &=&C_{\rm T2}+C_{\rm B2}+C_{\rm D}+C_{\rm S}+C_{\rm M}.  \notag
\end{eqnarray}

Then the capacitance matrix has the following form: 
\begin{equation}
\mathbf{C}=%
\begin{pmatrix}
C_{1} & -C_{\rm M} \\ 
-C_{\rm M} & C_{2}%
\end{pmatrix}%
.  \label{eq:capacitance}
\end{equation}%
Let us assume for simplicity that $V_{\rm S}=V_{\rm D}=0$. Then putting expressions
for the vector of charges \eqref{eq:charge_matrix} and for the inverse
capacitance matrix \eqref{eq:capacitance}, for the DQD electrostatic energy we obtain:

\begin{widetext}
\begin{eqnarray}
E &=&\frac{1}{C_{1}C_{2}-C_{\rm M}^{2}}\left[ \frac{1}{2}C_{1}Q_{2}^{2}+\frac{1}{%
2}C_{2}Q_{1}^{2}+C_{\rm M}Q_{1}Q_{2}\right] + \frac{V_{\rm TG}}{C_{1}C_{2}-C_{\rm M}^{2}}\left[ C_{\rm T1}\left(
C_{\rm M}Q_{2}+C_{2}Q_{1}\right) +C_{\rm T2}\left(
C_{1}Q_{2}+C_{\rm M}Q_{1}\right) \right] +  \notag \\
&&\frac{V_{\rm BG}}{C_{1}C_{2}-C_{\rm M}^{2}}\left[ C_{\rm B1}\left(
C_{\rm M}Q_{2}+C_{2}Q_{1}\right) +C_{\rm B2}\left(
C_{1}Q_{2}+C_{\rm M}Q_{1}\right) \right] + \frac{V_{\rm TG}^{2}}{C_{1}C_{2}-C_{\rm M}^{2}}\left[\frac{1}{2}C_{1}C_{\rm T2}^{2}+%
\frac{1}{2}C_{2}C_{\rm T1}^{2}+ C_{\rm T1}C_{\rm T2}C_{\rm M}\right] + \notag \\
&&\frac{V_{\rm BG}^{2}}{C_{1}C_{2}-C_{\rm M}^{2}}\left[ \frac{1}{2}C_{1}C_{\rm T2}^{2}+%
\frac{1}{2}C_{2}C_{\rm T1}^{2}+C_{\rm T1}C_{\rm T2}C_{\rm M}\right].
\label{eq:energy1}
\end{eqnarray}
\end{widetext}
To simplify Eq.~(\ref{eq:energy1}), let us introduce new values, $%
N_{i}=-\frac{Q_{i}}{\left\vert e\right\vert }$, for the number of electrons
in the $i$th quantum dot, and reduced top-gate $n_{t}$ and back-gate $n_{b}$
voltages

\begin{eqnarray}
n_{t} &=&\frac{C_{\rm T1}V_{\rm TG}}{\left\vert e\right\vert }=\frac{1}{1+a}\frac{%
C_{\rm T2}V_{\rm TG}}{\left\vert e\right\vert }, \\
n_{b} &=&\frac{C_{\rm B1}V_{\rm BG}}{\left\vert e\right\vert }=\frac{1}{1-a}\frac{%
C_{\rm B2}V_{\rm BG}}{\left\vert e\right\vert },  \label{eq:expr_nt}
\end{eqnarray}%
where $a$ is an asymmetry factor in the gate couplings. Assuming $%
C_{1}=C_{2}=mC_{\rm M}$ and rewriting the quantum dot charges as $%
Q_{1(2)}=-\left\vert e\right\vert N_{1(2)}$, then for the energy of the quantum dot, we have
\begin{widetext}
\begin{eqnarray}
\frac{E_{N_1,N_2}}{E_{\rm C}} &=&\frac{1}{2}N_{1}^{2}+\frac{1}{2}N_{2}^{2}+\frac{N_{1}N_{2}%
}{m} -n_{t}\left[ N_{1}+\frac{N_{2}}{m}+(1+a)\left( N_{2}+\frac{N_{1}}{m}%
\right) \right]  \notag \\
&&-n_{b}\left[ N_{1}+\frac{N_{2}}{m}+(1-a)\left( N_{2}+\frac{N_{1}}{m}%
\right) \right]  + n_{t}^{2}\left[ \frac{1}{2}+\frac{1}{2}(1+a)^{2}+\frac{1+a}{m}\right] 
\notag \\
&&+n_{b}^{2}\left[ \frac{1}{2}+\frac{1}{2}(1-a)^{2}+\frac{1-a}{m}\right] +n_{t}n_{b}\left[ 2(1+\frac{1}{m})-a^{2}\right],  
\label{eq:U}
\end{eqnarray}
\end{widetext}
where we defined $E_{\rm C}=E_{\rm C_1}=E_{\rm C_2}=mE_{\rm C_M}=e^{2}\frac{C_{1}}{{C_{1}^{2}-C_{\rm M}^{2}}}$. We plot the energy-level diagram in Fig.~\ref{energy_levels}(b) for the following parameters: $a=0.1$, $n_{b}=0.25$, $m=10$, and $N_{1},~N_{2}$ are equal to $0$ or $1$.

\section*{Data availability}
All data generated or analyzed during this study are included in this published article.

\nocite{apsrev41Control} 
\bibliographystyle{apsrev4-1}
\bibliography{references}

\end{document}